\newif\ifclean\cleantrue\newif\ifanon\anonfalse\newif\ifarxiv\arxivtrue

\newif\ifoldtodo
\oldtodofalse

\documentclass[acmsmall,nonacm,screen,timestamp]{acmart}

\usepackage[normalem]{ulem}

\usepackage{xcolor}

\hypersetup{colorlinks=true,allcolors=green}

\makeatletter
\NAT@numberstrue\NAT@supertrue
\renewcommand\NAT@open{}\renewcommand\NAT@close{}
\makeatother

\makeatletter
  \renewcommand\ACM@timestamp{%
    \footnotesize%
    \ifx\@acmSubmissionID\@empty\relax\else
    Submission ID: \@acmSubmissionID.{ }%
    \fi
    Draft of \the\year-\two@digits{\the\month}-\two@digits{\the\day}{ }%
    \two@digits{\theACM@time@hours}:\two@digits{\theACM@time@minutes}{. }%
    Page \thepage%
  }
\makeatother

\ifarxiv
\else
\usepackage{lineno}

\linenumbers
\fi

\usepackage{wrapfig}

\makeatletter
\newcommand{\mysection}[1]{
      \fancyhead[LO]{\ACM@linecountL\shorttitle: #1}%
\section{#1}
}
\newcommand{\mysubsection}[2]{
      \fancyhead[LO]{\ACM@linecountL\shorttitle: #1 (#2)}%
\subsection{#2}
}
\makeatother

\AtBeginDocument{%
  }

\newcommand{\mylink}[3]{\href{#3}{#1}}

\usepackage{tikz}

\usepgflibrary{arrows,decorations.markings}
\usetikzlibrary{shapes,graphs,quotes,arrows.meta,positioning,calc,chains,trees}
\usetikzlibrary{decorations,trees,spy,shadows,backgrounds,fit,trees,matrix,automata}

\definecolor{mynotgray}{RGB}{0,0,0}

\tikzset{font=\sffamily}
\newcommand{\myonly}[2]{#2}

\tikzset{pics/.cd,
magicjigsaw/.style args={#1/#2 and #3/#4/#5/#6 sz #7 col #8 text #9}{
    code={%
\draw[thick, white, fill=#8] (-#1,-0.35*#7) to[out=90,in={90+#3*45}] ({-#1+0.5*#3*#7},-0.45*#7)
arc({-135-(#3-1)*45}:{(#3-1)*180+135+(#3-1)*45}:{0.6*#7} and {0.45*sqrt(2)*#7})
to[out=-90-#3*45,in=-90] (-#1,0.35*#7) |- (-0.35*#7,#2)
to[out=0,in={0+#4*45}] (-0.45*#7,#2-0.5*#4*#7)
arc(180-#4*45:{(#4+1)*180+#4*45}:{0.45*sqrt(2)*#7} and 0.6*#7)
to[out=-180-#4*45,in=180] (0.35*#7,#2) -| (#1,0.35*#7)
to[out=-90,in=270+#5*45] (#1-#5*0.5*#7,0.45*#7)
arc(90-#5*45:{(#5+1)*180-90+#5*45}:0.6*#7 and {0.45*sqrt(2)*#7})
to[out=90-#5*45,in=90] (#1,-0.35*#7) |- (0.35*#7,-#2)
to[out=180,in=-180+#6*45] (0.45*#7,-#2+#6*0.5*#7)
arc(-#6*45:{(#6-1)*180+180+#6*45}:{0.45*sqrt(2)*#7} and 0.6*#7)
to[out=-#6*45,in=0] (-0.35*#7,-#2) -| cycle;
\node[align=center] at (0,0) {#9};
}}
}

\tikzset{pics/.cd,
Magicjigsaw/.style args={#1/#2 and #3/#4/#5/#6 sz #7 col #8 text #9}{
    code={%
\draw[thick, black, fill=#8] (-#1,-0.35*#7) to[out=90,in={90+#3*45}] ({-#1+0.5*#3*#7},-0.45*#7)
arc({-135-(#3-1)*45}:{(#3-1)*180+135+(#3-1)*45}:{0.6*#7} and {0.45*sqrt(2)*#7})
to[out=-90-#3*45,in=-90] (-#1,0.35*#7) |- (-0.35*#7,#2)
to[out=0,in={0+#4*45}] (-0.45*#7,#2-0.5*#4*#7)
arc(180-#4*45:{(#4+1)*180+#4*45}:{0.45*sqrt(2)*#7} and 0.6*#7)
to[out=-180-#4*45,in=180] (0.35*#7,#2) -| (#1,0.35*#7)
to[out=-90,in=270+#5*45] (#1-#5*0.5*#7,0.45*#7)
arc(90-#5*45:{(#5+1)*180-90+#5*45}:0.6*#7 and {0.45*sqrt(2)*#7})
to[out=90-#5*45,in=90] (#1,-0.35*#7) |- (0.35*#7,-#2)
to[out=180,in=-180+#6*45] (0.45*#7,-#2+#6*0.5*#7)
arc(-#6*45:{(#6-1)*180+180+#6*45}:{0.45*sqrt(2)*#7} and 0.6*#7)
to[out=-#6*45,in=0] (-0.35*#7,-#2) -| cycle;
\node[align=center] at (0,0) {#9};
}}
}

\newcommand{\myquitesmall}{\footnotesize}

\newcommand{\successNet}{{\makebox[0mm][r]{\raisebox{0mm}[0mm][0mm]{\hspace*{0mm}\myquitesmall$\left.\mbox{\begin{tabular}{l}
\textbf{Network specifications}\\
P4\cite{DBLP:conf/sigcomm/RuffyLKHTSDPSF23}, TCP/IP\cite{DBLP:conf/sigcomm/BishopFNSSW05,DBLP:journals/jacm/BishopFMNRSSW19}
\end{tabular}}\right\}$}}}\hspace*{0mm}}

\newcommand{\successPL}{\makebox[0mm][r]{\raisebox{1mm}[0mm][0mm]{\hspace*{0mm}\myquitesmall$\left.\mbox{\begin{tabular}{l}
\textbf{Programming language semantics}\\
C\cite{UCAM-CL-TR-981}, Rust\cite{DBLP:journals/pacmpl/JungKPMSW26}\\
\textbf{Programming language analysis and verification tools}\\
\mylink{Astr\'{e}e}{AstreeWWW}{https://www.absint.com/astree/},
\mylink{CBMC}{CBMCWWW}{https://www.cprover.org/cbmc/},
CN\cite{spec-test-prove-draft},
\mylink{Dafny}{DafnyWWW}{https://dafny.org/},
\mylink{Frama-C}{FramaCWWW}{https://frama-c.com/},
\mylink{Gillian}{GillianWWW}{https://vtss.doc.ic.ac.uk/research/gillian.html},
\mylink{Infer}{InferWWW}{https://fbinfer.com/},\\
\mylink{Iris}{IrisWWW}{https://iris-project.org/},
\mylink{KLEE}{KLEEWWW}{https://klee-se.org/},
\mylink{Soteria}{SoteriaWWW}{https://soteria-tools.com/},
\mylink{Velvet}{VelvetWWW}{https://github.com/verse-lab/velvet/},
\mylink{VeriFast}{VeriFastWWW}{https://github.com/verifast/verifast},
\mylink{Verus}{VerusWWW}{https://github.com/verus-lang/verus},
\mylink{Viper}{ViperWWW}{https://www.pm.inf.ethz.ch/research/viper.html},
\mylink{VST}{VSTWWW}{https://vst.cs.princeton.edu/}
\end{tabular}}\right\}$}}\hspace*{0mm}}

\newcommand{\successIR}{\makebox[0mm][r]{\raisebox{-2mm}[0mm][0mm]{\hspace*{0mm}\myquitesmall$\left.\mbox{\begin{tabular}{l}
\textbf{Intermediate languages}\\
LLVM: Alive2\cite{DBLP:conf/pldi/LopesLHLR21}, \mylink{Vellvm}{VellvmWWW}{https://vellvm.github.io/vellvm/}, MLIR\cite{DBLP:journals/pacmpl/FehrFPRG25}\\%
\myonly{4-}{WebAssembly\cite{DBLP:journals/pacmpl/YounSLRBGLPRWR24}}
\end{tabular}}\right\}$}}\hspace*{0mm}}

\newcommand{\successISA}{\makebox[0mm][r]{\hspace*{0mm}\myquitesmall$\left.\mbox{\begin{tabular}{l}
\textbf{Instruction-Set Architecture}\\
Arm\cite{DBLP:conf/fmcad/Reid16,DBLP:journals/pacmpl/ArmstrongBCRGNM19,ASL1www}\\
RISC-V\cite{sail-riscv}\\
CHERI-RISC-V\cite{UCAM-CL-TR-987}, CHERIoT\cite{cheriotmicro2023}, Morello\cite{DBLP:conf/esop/BauereissCSAESB22}
\end{tabular}}\right\}$}\hspace*{0mm}}

\newcommand{\successHWV}{\makebox[0mm][r]{\raisebox{-0mm}[0mm][0mm]{\hspace*{0mm}\myquitesmall$\left.\mbox{\begin{tabular}{l}
\textbf{Microarchitecture and h/w verif}\\
x86 in ACL2\cite{DBLP:conf/cpp/GoelSSS20}
\end{tabular}}\right\}$}}\hspace*{0mm}}

\newcommand{\successCloud}{{\makebox[0mm][l]{\raisebox{0mm}[0mm][0mm]{\hspace*{0mm}\myquitesmall$\left\{\mbox{\hspace*{-0mm}\begin{tabular}{l}
\textbf{Cloud crypto, parsing, etc.}\\ \mylink{F*}{FstarWWW}{https://fstar-lang.org}\\
\textbf{AWS infrastructure}\\ Amazon~\cite{DBLP:journals/queue/BrookerD24}
\end{tabular}}\right.$\hspace*{0mm}}}}}

\newcommand{\successHyp}{\makebox[0mm][l]{\raisebox{0.0mm}[0mm][0mm]{\hspace*{0mm}\myquitesmall$\left\{\mbox{\begin{tabular}{l}
\textbf{Hypervisors} SeL4\cite{KleinAEHCDEEKNSTW10}, Nitro\cite{nitroblog}, pKVM\cite{DBLP:conf/sosp/MemarianSKPS25}
\end{tabular}}\right.$}}\hspace*{0mm}}

\newcommand{\successRLXP}{\makebox[0mm][l]{\hspace*{0mm}\myquitesmall$\left\{\mbox{\hspace*{-0mm}\begin{tabular}{l}
\textbf{Relaxed Concurrency -- PL}\\
C/C++\cite{DBLP:conf/popl/BattyOSSW11}, JavaScript, Wasm
\end{tabular}}\right.$\hspace*{0mm}}}

\newcommand{\successCompCert}{\makebox[0mm][l]{\hspace*{0mm}\myquitesmall$\left\{\mbox{\begin{tabular}{l}
\textbf{Verified Compilation}\\
CompCert\cite{CompCertWWW}, CakeML\cite{CakeMLWWW}
\end{tabular}}\right.$\hspace*{0mm}}}

\newcommand{\successRLXA}{\makebox[0mm][l]{\hspace*{0mm}\myquitesmall$\left\{\mbox{\begin{tabular}{l}
\textbf{Relaxed Concurrency -- Arch}\\
x86\cite{DBLP:journals/cacm/SewellSONM10},
Arm\cite{DBLP:journals/pacmpl/PulteFDFSS18,DBLP:journals/toplas/AlglaveDGHM21,simner2025,DBLP:journals/pacmpl/PeramiBCLLAS26},
RISC-V\cite{riscv-unpriv-20191213}
\end{tabular}}\right.$\hspace*{0mm}}}

\newcommand{\successSoC}{\makebox[0mm][l]{\raisebox{-0mm}[0mm][0mm]{\hspace*{0mm}\myquitesmall$\left\{\mbox{\begin{tabular}{l}
\textbf{SoC semantics}\\
Sockeye\cite{fiedler2026sockeyelanguageanalyzinghardware}
\end{tabular}}\right.$}}\hspace*{0mm}}

\newcommand{\cartoonCloud}{  \draw (5.0,6.6)  node (ejig) {} pic{magicjigsaw=2.5/0.5 and 0/0/0/0  sz 0.6 col tests text {\huge Cloud Services}};}
\newcommand{\cartoonNetwork}{\draw (5.0,5.6)  node (ejig) {} pic{magicjigsaw=2.5/0.5 and 0/0/0/0  sz 0.6 col tests text {\huge Network}};}
\newcommand{\cartoonOS}{     \draw (5.0,4.6)  node (ejig) {} pic{magicjigsaw=2.5/0.5 and 0/0/0/0  sz 0.6 col tests text {\huge OS, Hypervisor}};}

\newcommand{\cartoonCompCla}{\draw (4.5,3.35) node (ejig) {} pic{magicjigsaw=1.1/0.75 and 0/0/0/0  sz 0.6 col tests text {\LARGE Clang}};}
\newcommand{\cartoonCompNCl}{\draw (5.8,3.35) node (ejig) {} pic{magicjigsaw=0.3/0.75 and 0/0/0/0  sz 0.6 col tests text {\huge }};}
\newcommand{\cartoonCompLLV}{\draw (4.3,1.85) node (ejig) {} pic{magicjigsaw=0.8/0.75 and 0/0/0/0  sz 0.6 col tests text {\Large LLVM}};}
\newcommand{\cartoonCompWas}{\draw (5.5,1.85) node (ejig) {} pic{magicjigsaw=0.6/0.75 and 0/0/0/0  sz 0.6 col tests text {Wasm}};}
\newcommand{\cartoonCompGCC}{\draw (6.8,2.6)  node (ejig) {} pic{magicjigsaw=0.7/1.5 and 0/0/0/0  sz 0.6 col tests text {\LARGE GCC}};}
\newcommand{\CartoonCompilers}{\draw (3.0,2.6) node (ejig) {} pic{magicjigsaw=0.5/1.5  and 0/0/0/0  sz 0.6 col tests!100 text {\rotatebox[origin=c]{270}{\huge{}Compilers}{\myonly{10-}{\color{mynotgray}}%
}}};}

\newcommand{\cartoonHW}{     \draw (5.0,0.1)  node (ejig) {} pic{magicjigsaw=2.5/1.0 and 0/0/0/0  sz 0.6 col tests text {\huge Hardware\\\Large Processors, SoCs}};}

\newcommand{\CartoonCloud}{  \draw (5.0,6.6)  node (pjig) {} pic{magicjigsaw=2.5/0.5  and 0/0/0/0  sz 0.6 col tests!50 text {
      Cloud Infra
}};}
\newcommand{\CartoonNetwork}{\draw (5.0,5.6)  node (ejig) {} pic{magicjigsaw=2.5/0.5  and 0/0/0/0  sz 0.6 col tests!50 text {Network  }};}
\newcommand{\CartoonOS}{     \draw (5.0,4.6)  node (ejig) {} pic{magicjigsaw=2.5/0.5  and 0/0/0/0  sz 0.6 col tests!50 text {OS, Hypervisors}};}
\newcommand{\CartoonSeLFour}{\draw (7.1,4.6)  node (ejig) {} pic{magicjigsaw=0.4/0.5  and 0/0/0/0  sz 0.6 col tests!100 text {SeL4}};}

\newcommand{\CartoonCompCla}{\draw (3.7,3.35) node (ejig) {} pic{magicjigsaw=1.2/0.75 and 0/0/0/0  sz 0.6 col tests!50 text {Clang}};}
\newcommand{\CartoonCompNCl}{\draw (5.2,3.35) node (ejig) {} pic{magicjigsaw=0.3/0.75 and 0/0/0/0  sz 0.6 col tests!50 text {\huge }};}
\newcommand{\CartoonCompLLV}{\draw (3.4,1.85) node (ejig) {} pic{magicjigsaw=0.9/0.75 and 0/0/0/0  sz 0.6 col tests!50 text {LLVM}};}
\newcommand{\CartoonCompWas}{\draw (4.9,1.85) node (ejig) {} pic{magicjigsaw=0.6/0.75 and 0/0/0/0  sz 0.6 col tests!50 text {Wasm}};}
\newcommand{\CartoonCompGCC}{\draw (6.5,2.6)  node (ejig) {} pic{magicjigsaw=1.0/1.5  and 0/0/0/0  sz 0.6 col tests!50 text { GCC}};}
\newcommand{\CartoonCompCert}{\draw (7.3,2.6) node (ejig) {} pic{magicjigsaw=0.2/1.5  and 0/0/0/0  sz 0.6 col tests!100 text {\rotatebox[origin=c]{270}{\tiny{}CompCert,CakeML}}};
\draw (8.0,2.6) node {\successCompCert};}
\newcommand{\CartoonHW}{     \draw (5.0,0.1)  node (ejig) {} pic{magicjigsaw=2.5/1.0  and 0/0/0/0  sz 0.6 col tests!50 text {Hardware}};}

\newcommand{\cartoonInterfaces}{
\myonly{1-}{\draw (5.0,-0.5) node (ejig) {} pic{Magicjigsaw=2.7/0.2 and 0/0/0/0  sz 0.4 col tests text {\normalsize MicroOps Spec}};}
\myonly{1-}{\draw (2.0,-0.5) node {\successHWV};}
\myonly{1-}{\draw (8.0,-0.2) node {\successSoC};}
\myonly{2-}{\draw (5.0,0.95) node (ejig) {} pic{Magicjigsaw=2.7/0.4 and 0/0/0/0  sz 0.6 col tests text {\large Architecture Spec}};}
\myonly{2-}{\draw (2.0,0.95) node {\successISA};}
\myonly{2-}{\draw (8.0,0.95) node {\successRLXA};}
\myonly{3-}{\draw (3.3,2.55) node (ejig) {} pic{Magicjigsaw=1.0/0.2 and 0/0/0/0  sz 0.6 col tests text {\normalsize LLVM}};}
\myonly{3-}{\draw (2.0,2.55) node {\successIR};}
\myonly{4-}{\draw (4.9,2.55) node (ejig) {} pic{Magicjigsaw=0.6/0.2 and 0/0/0/0  sz 0.6 col tests text {\normalsize Wasm}};}
\myonly{5-}{\draw (5.0,3.95) node (ejig) {} pic{Magicjigsaw=2.7/0.4 and 0/0/0/0  sz 0.6 col tests text {\large Programming Language Spec}};}
\myonly{5-}{\draw (2.0,3.95) node {\successPL};}
\myonly{5-}{\draw (8.0,3.95) node {\successRLXP};}
\myonly{7-}{\draw (5.0,5.1)  node (ejig) {} pic{Magicjigsaw=2.7/0.3 and 0/0/0/0  sz 0.6 col tests text {\large OS API Spec}};}
\myonly{7-}{\draw (8.0,5.1)  node {\successHyp};}
\myonly{8-}{\draw (5.0,6.1)  node (ejig) {} pic{Magicjigsaw=2.7/0.3 and 0/0/0/0  sz 0.6 col tests text {\large Network Protocol\,\&\,Lang Spec}};}
\myonly{8-}{\draw (2.0,6.1)  node {\successNet};}
\myonly{8-}{\draw (8.0,6.1)  node {\successCloud};}
}

\newcommand\myparagraph[1]{\noindentparagraph{\bfseries\upshape{#1}}}
\newcommand\mysubparagraph[1]{\noindentparagraph{\bfseries\itshape{#1}}}
\usepackage{proof}
\usepackage{enumitem}

\definecolor{tests}{HTML}{F9DD25}
\definecolor{ibmcolourblind2}{RGB}{120, 94, 240} %
\definecolor{ibmcolourblind3}{RGB}{220, 38, 127} %

\ifclean
\newcommand\TODO[1]{}
\newcommand\TODOJP[1]{}
\newcommand\TODOPSSUPPRESS[1]{}
\newcommand\TODOPS[1]{}
\else
\newcommand\TODO[1]{{\color{ibmcolourblind1}TODO: #1}}
\newcommand\TODOJP[1]{{\color{ibmcolourblind3}JP: #1}}
\newcommand\TODOPSSUPPRESS[1]{{\color{ibmcolourblind2!40}PS: #1}}
\newcommand\TODOPS[1]{{\color{ibmcolourblind2}PS: #1}}
\fi

\ifoldtodo
\newcommand{\OLD}[1]{\marginpar{\tiny{}#1}}
\else
\newcommand{\OLD}[1]{}
\fi

\begin{document}

\title[Escaping the Quicksand: a Call to Arms]{Escaping the Quicksand: A Call to Arms}

\author{Peter Sewell}
\affiliation{
  \institution{University of Cambridge}
  \country{UK}
}
\email{Peter.Sewell@cl.cam.ac.uk}

\author{Jean Pichon-Pharabod}
\affiliation{
  \institution{Aarhus University}
  \country{Denmark}
}
\email{jean.pichon@cs.au.dk}

\begin{CCSXML}
<ccs2012>
<concept>
<concept_id>10011007</concept_id>
<concept_desc>Software and its engineering</concept_desc>
<concept_significance>500</concept_significance>
</concept>
<concept>
<concept_id>10003752.10010124</concept_id>
<concept_desc>Theory of computation~Semantics and reasoning</concept_desc>
<concept_significance>500</concept_significance>
</concept>
</ccs2012>
\end{CCSXML}

\ccsdesc[500]{Software and its engineering}
\ccsdesc[500]{Theory of computation~Semantics and reasoning}

\maketitle

{\em
Computing has been an astonishing success -- but the accumulated technical debt exposes us all to huge costs in business and societal risk.  For 75 years, we've built systems to prose specifications with test-and-debug development. We made that work well enough for the industry to thrive, but it's an expensive and ineffective feedback loop,
that's left everyone depending on shaky foundations. 
Now, AI-enabled engineering is amplifying the success by reducing coding costs, but also amplifying the risks:
AI test-and-debug development is rapidly increasing our technical debt,
and AI vulnerability detection is making it ever-easier to exploit that debt.

How can we do better?  Much research has pursued mathematical proof of correctness, which, unlike testing, can cover all cases.  This too has advanced massively, but it remains infeasible for normal practice, both for technical reasons and because of a deep-seated cultural disconnect, between those focussed on testing and on proof.

Instead, we argue for a pragmatic approach: flexible combinations of testing, \textbf{specification}, and proof, that can provide new and more effective feedback loops for both AI and human development, in any combination.

Most simply, with existing tools, 
one can incrementally co-develop executable-as-test-oracle
partial 
specifications alongside conventional prose descriptions, code, and tests, expressing them in the ambient programming language and tensioning against the code just by testing. This clarifies design
and gives a much tighter feedback loop, without needing exotic tools or skills.  It could be
code-first, retrofitting specs to existing code; spec-first,
writing or synthesising executable specs that
constrain 
code synthesis;
or any hybrid.
Developers
can and should do this today.

Or, even better,
one can use specifications that support the full gamut of testing, property-based testing, symbolic execution, and proof.
This enables a range of intertwined feedback loops, again both for AI and human, from cheap testing through to more expensive proof:
improving
specifications and code first by testing in concrete execution, then by property-based testing and symbolic execution, and then proof. 
It gives a gentle on-ramp to gradually increasing assurance,
with benefits commensurate to effort.
However, making it really practical needs a body of \textbf{semantics infrastructure}: specifications and tooling for the main programming languages and other abstractions, which we now more-or-less know how to build, but which is not yet in place. We call the community to arms to create and deploy it -- to enable a future built on
firmer ground.

}

\myparagraph{Computing has been an astonishing success -- but at a huge cost in business and societal risk}

Computing has
transformed the world since its origins in the 1930s and 1940s. 
Advances in computer science and engineering have given us systems of remarkable performance, sophistication, and impact, building
computing infrastructure that is now fundamental to modern society, across personal, economic, medical,
and government spheres.

But it comes at a cost. There is the obvious development cost:
building new systems to a sufficient quality to be marketable is expensive.
More importantly, we pay hidden costs in technical debt and risk.
Our infrastructure --  the millions of lines of code that implement everything from our hardware designs to phones to cloud services --
works well enough in normal use to be marketable,
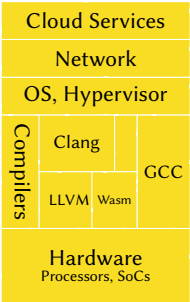
\begin{wrapfigure}{r}{0.25\textwidth}
\vspace*{-1.0\baselineskip}
\begin{center}
\noindent\hspace*{-0mm}
\scalebox{0.5}{
\begin{tikzpicture}
 \draw[thick, white] 
    (5.0,8.0);
{\cartoonHW}
{
\cartoonOS
\cartoonNetwork
\cartoonCloud
}
{
\cartoonCompCla
\cartoonCompNCl
\cartoonCompLLV
\cartoonCompWas
\cartoonCompGCC
\CartoonCompilers
}
\end{tikzpicture}
}%
\end{center}
\caption{Parts of the computing infrastructure that we all rely on}
\end{wrapfigure}
but it is riddled with flaws that all too often can be exploited, and, in an increasingly adversarial environment, all too often are.
We have built systems that are far too complex to be fully understood, on a quicksand foundation of decades of legacy choices.
This
exposes us all to continual risk of malfunction and continual to risk of attack, at every scale from individual criminals to nation states.
Imagine, if you don't want to sleep tonight, how long an outage of our computing and communications infrastructure a business, hospital, or society as a whole could withstand without collapse.
And, while there always have been outages due to simple failures and malicious attackers, now attacks can occur even without human intent, as we try to encapsulate AI -- trained to find security vulnerabilities -- using just the flawed mechanisms developed over previous decades\cite{huggingpost}.

Current attempts to mitigate such risks, in technical,
social, and legal ways, are vital, but they are also a continual Red Queen's race that can never be won.

\myparagraph{How did we get ourselves into this? By relying on test-and-debug development, focussing solely on code while neglecting specification, %
and 
misaligning market and societal incentives.}
Conventional engineering relies on prose specification and on test-and-debug development.
We build systems from millions of lines of code, without precise descriptions of what it should achieve, or the reasons why it is correct.
And we develop them by testing on concrete inputs,
checking the results, and debugging and fixing any issues that reveals.

This is an expensive and ineffective feedback loop. 
Specifications are used to communicate
between human beings, and now also in prose prompts for LLMs. Prose does make them superficially accessible,
but prose is intrinsically
a poor medium for describing the subtle and complex behaviour of real systems. Prose descriptions
are almost inevitably ambiguous and incomplete; they are often in some ways inconsistent or
simply incomprehensible; and they do not directly support any rigorous mechanised use -- one cannot directly test against them, let alone use them for test
generation, or coverage analysis, or generation of implementation components, or static analysis,
or proof.
Prose specifications can thus only be very weakly tensioned against implementation, by
curating intended test-case results from the prose, falling back to
implicit specifications such as the absence of crashes,
or relying on opaque AI code analysis. 

Then testing alone can cover only a tiny fraction of the more-than-astronomical number of possible inputs, execution paths, and internal states of real systems: for a deterministic program that takes a megabyte of input, there are around $10^{1\,000\,000}$ possible inputs, compared with the mere $10^{80}$ atoms in the observable universe.
Conventional testing manifestly can suffice to make systems that work well enough in common cases to market and use, but it manifestly cannot make systems robust or secure enough for today’s adversarial environment.

Exacerbating the challenge is the scale and pan-industry nature of the fundamental components: the major processor architectures, ABIs (application binary interfaces), programming languages, compilers, network protocols, operating-system APIs, and so on.
We have become locked into 
vast technical debt, based on design decisions taken in the 1960s
and 1970s in a much less adversarial environment. Then, computing resources were scarce,
we relied on them much less,
programming
languages and other protection technologies were much less developed, and malicious cyber-attack was almost
unknown. The resulting designs were arguably good choices then, but %
they remain the basis for our infrastructure today: hardware provides protection only at coarse granularities, with virtual memory and privileged
execution modes, and systems software remains largely written in C and C++.
Recent years have seen a welcome focus on software and hardware memory safety, e.g.\ with Rust and CHERI~\cite{DBLP:journals/cacm/WatsonBCCCDFGJLMMNORRST25}, but the deployed base is so large that wholesale replacement is infeasible. Fixing memory safety would be a big step forwards, but even if we can, higher- and lower-level problems will quickly come to the fore.

These technical challenges are intertwined with mismatched incentives and a market failure. Vendors have huge incentives to bring new systems quickly to market, and thus to use the established infrastructure and engineering skills base, but that locks us in still further, while the risks fall largely on the end-users and on society as a whole.
The option of reduced risk is usually not even available, but where it is, if it appears principally as a cost, then organisations often choose not to pay it.

\myparagraph{AI will save us, right?}
AI-enabled engineering promises to amplify all of this, both the success and the costs and risks. 
AI coding promises to reduce coding costs, 
AI bug-finding finds many security vulnerabilities~\cite{chromerace},
and
AI translation might reduce our dependence on legacy languages.
All are very attractive, but together they seem much more likely to accelerate the Red Queen's race than to end it.
Cheaper coding will lead to ever-larger codebases, and AI coding currently relies on the same ineffective test-and-debug feedback loop: ever-larger codebases will mean ever more opportunities for exploitable errors.
AI bug-finding can catch some of these, but we have no reason to believe that it will catch all. The rapid pace of automated vulnerability detection serves the attacker just as well as (or even better than) the defender, and will place automated discovery of zero-days in the hands of many.
Meanwhile, dependence on all these will mean that the human knowledge of the codebase, and the human skills to understand, engineer, and debug it, will atrophy: when the AI engineering is not up to the task, we will have no alternative.
Above we argued that human engineering needs better feedback loops -- but AI-based engineering needs them even more acutely.

\myparagraph{Formal mathematics will save us, right?}
A long history of work, also from the 1930s--40s onwards, explores more rigorous development methods:
mathematical specification of the desired behaviour of computer systems,
more cleanly structured and less error-prone programming languages and other abstractions, and
more discriminating ways to check that implementations have the desired properties, with richer type systems, static analysis, and proof.

For many decades, despite consistent advances and some impressive
results, this struggled to gain widespread industry traction.
Some ideas fed into practice, but mainstream engineering continued to
rely on test-and-debug development, prose specifications (or none
at all), and legacy languages.
Our abilities to specify and verify
systems improved massively, but were outstripped by the huge increases
in system complexity; and
there was a cultural disconnect between 
the classic formal methods and semantics community and mainstream practice -- some of the former (from Dijkstra onwards) emphasising formal approaches and proof \emph{in opposition} to testing, requiring a radical shift to mainstream practice and quite different skills, rather than methods that complement and smoothly extend existing practice.
The lack of approaches and tools that could be directly applied to mainstream development,
the weight of legacy systems and skills,
and the incentive structure emphasising rapid development and viewing improvements in engineering methods as a cost rather than a benefit,
combined to impede improvement.

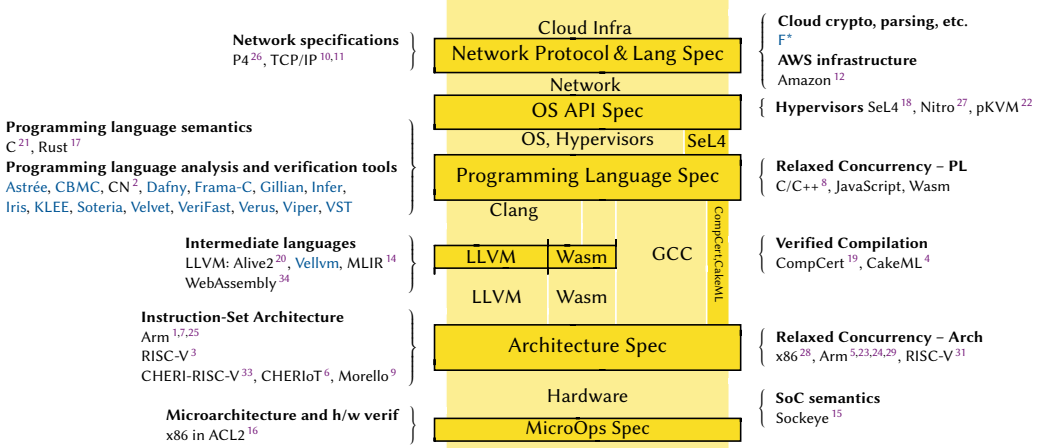
\begin{figure}[h]
\hspace*{20mm}\scalebox{0.75}{
\begin{tikzpicture}
\CartoonHW
\CartoonOS
\CartoonNetwork
\CartoonCloud

\CartoonCompCla
\CartoonCompNCl
\CartoonCompLLV
\CartoonCompWas
\CartoonCompGCC
{\CartoonSeLFour}
{\CartoonCompCert}

\cartoonInterfaces

\end{tikzpicture}
}%
\caption{Selected semantics success stories
}\label{figsuccess}
\end{figure}

\myparagraph{Pragmatic semantics, specification, and verification}
More recently, it has become clear that it is possible to do
better, at scale, in pragmatic ways that can be more smoothly adopted into and impact mainstream engineering. %
Fig.~\ref{figsuccess} shows some recent success stories, rigorously defining abstractions at various levels in the stack, and using those to improve engineering -- both in lightweight ways and in some cases with full  correctness proofs, of the implementation of one interface above another.  In this short paper we can mention just a sample, so apologies to those omitted -- this is not remotely a complete survey. The 2024 \href{https://www.newton.ac.uk/event/bsp/}{Big Specification programme} at the Isaac Newton Institute included talks on many of these.

Several advances have come together to permit this.  Some are technical: advances in tools for mechanised mathematics (ACL2, CVC5, HOL4, Isabelle, Lean, Rocq, Z3, etc.), in logics, analysis methods, and tools for reasoning about programs,
and in greater computing capacity.
But a key advance is simply a change of mindset: focussing on rigorous specification, on the feedback loops that can enable, and on methods that can be applied to existing systems not just clean-slate redesigns.
Note that we are advocating a \emph{focus} on specification in many forms and for many uses. We are \emph{not} arguing that one should normally write formal specifications up-front, but rather that one needs a nuanced understanding of the best possible workflows in different contexts.

\mysubparagraph{Making specifications executable as test oracles}
One of the simplest, and perhaps the most under-appreciated, use of more rigorous specifications
is to use them as test oracles. We say a specification is \emph{executable as a test oracle} if, given some
behaviour that might be exhibited by the system, the specification can be used to compute whether
or not that behaviour is allowed\cite{DBLP:conf/sigcomm/BishopFNSSW05}.

Given an executable-as-test-oracle specification, and instrumentation to record the externally
observable behaviours of an implementation, one can mechanically test whether the specification
and an implementation are consistent, simply by running whatever tests one has and checking
whether the specification allows the observed behaviours. This might be to check whether the
implementation correctly implements the specification, or to validate whether the specification
soundly abstracts from the implementation, or both. It makes the specification a live engineering
artifact, that can be integrated into conventional development practice.

It also radically simplifies the construction of test suites: given such a specification, one can
make tests automatically (systematically or randomly),
without needing a manually written check or manually curated set of allowed results for every
test.

A specification that is executable as a test oracle can take many forms. It could simply be
executable assertions, or executable pre- and post-conditions or other contracts, or a program that
takes a representation of a system behaviour (perhaps a trace) as input and checks whether it has
some desired properties, or, in some cases, a reference implementation. It has to be mechanised
in some way, but this need not be in an exotic specification language. Any of the above could be
expressed in a conventional programming language -- either the ambient language of the implementation, or a functional language chosen for its clarity --  or in some mechanised mathematical
form, equipped with some execution mechanism.

\mysubparagraph{Executable and testable abstraction functions and invariants}
Specification and specification testing can be just for the top-level behaviour of some component,
but it can also be much more fine-grained: one can define computable abstraction functions, that compute the intended abstract state from the current concrete implementation state, and check these step-by-step\cite{DBLP:conf/sosp/MemarianSKPS25}.  This can be very discriminating, capturing and checking -- even without proof -- many of the reasons why a system behaves as intended.  If one can examine and diff the abstract states, it also gives a powerful way to inspect a complex system.

\mysubparagraph{Partial specification and a gentle on-ramp to full verification}
Specifications that support both testing and proof offer an approach to incrementally increasing assurance: improving both
code and specifications with feedback from testing and property-based testing, which check that
the code and specification are consistent in concrete executions, and then (if more assurance
is desired) attempting proof\cite{spec-test-prove-draft}. This is a simple and obvious idea that dates back to the 1970s and should have become routine
practice long ago, but somehow has not. A long history of lightweight
formal methods has advocated aspects of it and put some of the pieces in place, but it has not been
part of main-line thinking.
It could be
done either code-first, incrementally adding partial specifications for existing code; %
or specification-first, as in some classic formal-methods approaches; or co-developing both code
and specification. It could be done either
to provide strong feedback loops for AI tooling, or where both code and specification are human-written.
It provides a gentle on-ramp to full verification, with incremental benefits (and incremental demands on costs and skills) as one goes, rather than the more all-or-nothing character of traditional proof-focussed formal methods.

\mysubparagraph{Design clarification}
Perhaps most importantly, writing a precise specification, tensioned against implementation,
 can help clarify the intended abstraction. It is all too
easy to add a paragraph to a prose specification without considering all of its implications, and
very hard to check that a prose specification is self-consistent, or that it covers all that it should. In
contrast, when writing a more rigorous specification, one is forced to consider what it says in all
cases. When coupled with any serious use of the specification, by testing or proof, this tends to
expose open questions in the intent.
We and others have found this in multiple cases: 
the relaxed-memory concurrency of architectures and programming languages,
the semantics of C,
the aliasing and provenance semantics of C, LLVM IR, and Rust,
the undef and poison semantics of LLVM IR,
and even the definition of PDF files.
In each case the historical lack of precise specification has led to widespread misunderstandings and deep-seated mismatches baked into our infrastructure, which are expensive or even infeasible to fix up after the fact. 
For example, the `middle-end' of GCC assumes that the provenance of a pointer value matters, while the back-end assumes they are just machine integers -- which are fundamentally incompatible choices of what optimisations are permitted.
And no high-performance high-level language has a satisfactory definition of the allowed concurrent behaviour of ``relaxed-atomic'' racy programs.

\myparagraph{Effective feedback loops for humans and AI}
All this suggests that more effective and efficient feedback loops for computer engineering are possible,
for any combination of human and AI-driven development:
much more discriminating testing loops, to reduce error and to improve assurance that specifications are as intended;
and incremental development of specifications and code (either together or in either order) as a smooth on-ramp towards
fine-grained specifications that support mechanised proofs of correctness.
Mechanised proof can leverage the rapid advances in AI proof -- and mechanised proofs are checkable by small non-AI prover kernels, to provide actual assurance even for AI-coded systems.

\myparagraph{That sounds good -- so what's missing?}
Some of what we advocate can be done immediately, with existing tools.
For example, instead of conventional coding, or vibe-coding an implementation from a prose prompt, one can write -- or in some cases generate -- a rigorous specification in the ambient programming language, and generate both code and an executable abstraction function. Continuously testing that the code and specification are related by the abstraction function helps a lot to keep AI (or human) coding on the straight and narrow.
We need to \textbf{advocate} and \textbf{educate} developers about such techniques right now.  %

But much requires a body of \emph{semantics infrastructure}: well-validated and usable semantics of the abstractions on which the implemented system rests -- be they assembly, C, Rust, OS APIs, or whatever -- and testing/analysis/proof tooling for them.
These are the generic foundations, structural frame, and tools that are need to improve engineering and assurance for any system.
For example, if one wants AI proof (or human proof) of C code, one needs a well-validated semantics of C, a specification language, and robust tooling for testing, test generation, and proof that the code and specification are consistent -- as we've prototyped\cite{spec-test-prove-draft}.
Or if one wants proof of Rust systems code that manages hardware-architecture security features such as virtual memory, down to the binary -- e.g.\ a proof that a hypervisor securely encapsulates untrusted AI components -- one needs well-validated semantics of the underlying architecture, of the LLVM IR intermediate language, of linking and loading, and of the Rust static and dynamic semantics, and, again, tooling for testing, test generation, and proof that lets one relate each layer.
We arguably should have developed this over the last 50 years, in sync with the evolving architectures, programming languages, compilers, protocols, etc., but we did not.  For example, there is no specification language for the behaviour of C code that is remotely as well-established as C itself.
The success stories in Fig.~\ref{figsuccess} provide some of what's needed, and they demonstrate convincingly that it's possible to build it, even for some of the gnarliest of legacy abstractions -- it is possible to catch up.  But each has demanded a major project by a substantial group of researchers over many years.

\medskip

This is a challenge of scale, incentives, and market failure:

\medskip

The normal \textbf{academic incentives and funding models} are at odds with the demands of such projects. They need long-term investment and commitment, to develop semantic artifacts at a larger scale, engineer them and the associated tooling to make them usable in production, and maintain them into the indefinite future.  

The normal \textbf{industry incentives} are at odds with the fact that we depend on many \emph{pan-industry} abstractions -- for example, no one vendor controls or stands to exclusively benefit from improvements to the C specification, even though many would (directly or indirectly).  A well-resourced company might build parts of what is needed for their own use, and some have, but that is not enough. 
Then, while building the semantic infrastructure and tools requires broad investment, we have to ensure that using them provides benefits in proportion to commitment, in both engineering cost and system quality.  

It needs \textbf{community consensus} on how to do it, to make integrated solutions.  Only when one tries to integrate and use specifications does one discover how they should be phrased to make them actually usable, and only when one tries to relate multiple abstractions -- e.g.\ to prove correctness of compilation from Rust via LLVM to assembly and binary code -- can one understand how best to set up all the definitions to make that possible.  Isolated projects tackling different abstractions in different ways will not lead to semantic components that can be integrated later.

We now know much of how to do this, but many \textbf{research questions} remain. For example, ongoing work\cite{DBLP:journals/pacmpl/PeramiBCLLAS26} has integrated architecture-level instruction-set\cite{DBLP:conf/fmcad/Reid16,DBLP:journals/pacmpl/ArmstrongBCRGNM19,sail-riscv,ASL1www} and concurrency\cite{DBLP:journals/pacmpl/PulteFDFSS18,DBLP:journals/toplas/AlglaveDGHM21,simner2025} semantics, and considered the semantics (including aliasing and provenance models) of C\cite{UCAM-CL-TR-981,DBLP:conf/popl/BattyOSSW11}, Rust\cite{DBLP:journals/pacmpl/JungKPMSW26}, and LLVM IR\cite{DBLP:conf/pldi/LopesLHLR21,VellvmWWW,DBLP:journals/pacmpl/FehrFPRG25}, but integrating and reasoning about the combination of all these is an active topic, 
as is the design of really effective \emph{tool support for semantics}.

It needs \textbf{education} of undergraduates, PhD students, postdocs, and industry practitioners, both of those who can develop this semantic infrastructure (a relatively skilled activity) and of those who need to use it in day-to-day development.
Programming is widely taught, but where are the courses on specification, specification-based testing, and the whole gamut of rigorous engineering? 

And it needs \textbf{care and support to maintain the research culture}. Both AI and semantics and verification are currently facing a success disaster:
skilled researchers are in such high demand from well-resourced companies
that it is becoming impossible to retain the next generation of researchers, to advance the fundamentals and educate the future.
Traditionally limited academic salaries cannot compete with industry offers at many times their levels, for intellectually interesting and challenging work.

\medskip

Other disciplines realised long ago that they needed to support ``big science'',  for major biology, physics, or engineering research -- e.g.\ the human genome project, or particle physics, or aeronautics, but computer science has not.
The semantics infrastructure we need demands large resources by the standards of any one academic or industry research team -- but only small resources by the standards of industry as a whole, and an infinitesimal fraction of the current AI spend. 
It is larger than (say) a typical ERC call, or a UK ARIA or UKRI programme, or a DARPA programme, or normal academic funding from the large vendors (Amazon, AMD, Apple, Arm, IBM, Intel, Google, Meta, NVIDIA, etc.), or typical philanthropic funding,
but it does not need another CERN: \emph{fairly big science} would suffice to make a big step forwards: coordination of some or all of those potential funders and users, new social structures among researchers, and 100s of M\$, not of B\$, of funding. 

\medskip

So, this is a call to arms, to the research community and all those organisations, to collectively organise and make this happen: to build solid foundations, not quicksand, for future computing.

\begin{acks}
We thank all our academic and industry colleagues.
We thank Nick Benton, Sarah de Haas, Kathleen Fisher, Richard Grisenthwaite, and Warren Hunt for comments on drafts.
This work was funded in part by UK Research and Innovation (UKRI) under the UK government's Horizon Europe funding guarantee for ERC-AdG-2022, EP/Y035976/1 SAFER (Sewell).
This work was funded in part by an AUFF starter grant (Pichon-Pharabod).
We thank the Isaac Newton Institute for Mathematical Sciences, Cambridge, for support and hospitality during the Big Specification programme, which gave us the spur to write this piece, and thank our co-organisers and the other attendees.  This work was supported in part by EPSRC grant EP/Z000580/1.
\end{acks}

\bibliographystyle{ACM-Reference-Format}
\bibliography{refs-other,refs-sewell-dblp-185,refs-sewell-more}

\end{document}